\documentclass[11pt,draftcls,onecolumn]{IEEEtran}
\usepackage{pifont}
\usepackage{mathrsfs}
\usepackage{amssymb}
\usepackage{amsmath}
\usepackage{txfonts}
\usepackage{balance}
\usepackage{amsfonts}
\usepackage{color}
\usepackage{dsfont}
\usepackage{cite}
\usepackage{subfigure}
\usepackage[]{graphicx}
\ifCLASSINFOpdf
\else
\fi
\begin{document}
%
\title{Distributed Source Localization in Wireless
Underground Sensor Networks}

\author{Hongyang Chen, Robin Wentao Ouyang, and Chen Wang \thanks{

H. Chen is with the Institute of Industrial Science, the University
of Tokyo (e-mail: hongyang@mcl.iis.u-tokyo.ac.jp). R. W. Ouyang is
with the Department of Electronic and Computer Engineering, Hong
Kong University of Science and Technology, Hong Kong (email:
oywtece@ust.hk). C. Wang is with the Ministry of Education Key Lab
for Intelligent Networks and Network Security, Xi'an Jiaotong
University, Xi'an, China (e-mail: wangchen@ieee.org).


}}

\maketitle

\begin{abstract}
Node localization plays an important role in many practical
applications of wireless underground sensor networks (WUSNs), such
as finding the locations of earthquake epicenters, underground
explosions, and microseismic events in mines. It is more difficult to
obtain the time-difference-of-arrival (TDOA) measurements in WUSNs
than in terrestrial wireless sensor networks because of the unfavorable
channel characteristics in the underground environment. The robust
Chinese remainder theorem (RCRT) has been shown to be an effective
tool for solving the phase ambiguity problem and frequency estimation
problem in wireless sensor networks. In this paper, the RCRT is used to robustly
estimate TDOA or range difference in WUSNs and therefore improves the ranging
accuracy in such networks. After obtaining the range difference, distributed source
localization algorithms based on a diffusion strategy are proposed to
decrease the communication cost while satisfying the localization
accuracy requirement. Simulation results confirm the validity and
efficiency of the proposed methods.

\end{abstract}



%
\IEEEpeerreviewmaketitle

%
%

\section{Introduction}
Wireless underground sensor networks (WUSNs) are an important
extension of terrestrial wireless sensor networks, as they can be
used to estimate the location of earthquake epicenters, underground
explosions, microseismic activities in mines, etc. Normally, the
sensor nodes in WUSNs are buried underground and they exchange
information wirelessly via the dispersive underground channel. Some
experimental results with WUSNs are reported in \cite{4}.

In terrestrial wireless sensor networks, time difference of arrival
(TDOA) measurements are widely used for node localization \cite{1}.
Because of the physical characteristics of the dispersive
underground channel and the heterogeneous network architecture of
WUSNs, source localization for WUSNs based on TDOA is more
challenging \cite{wusn}. In a dispersive medium, we cannot directly
obtain the range differences between the source and sensors from
TDOA measurements since the propagation velocity is a function of
frequency;  different frequency components will have different
propagation delays \cite{2}.

Determining  the location of sensor nodes is important in many
practical applications of wireless underground sensor networks. The
objective of a positioning system is to determine accurate node
locations with low complexity and communication cost. Localization
algorithms for traditional terrestrial wireless sensor networks can
be classified into two types: range-based methods and range-free
methods. Range-based methods usually have higher location accuracy
than range-free ones while demanding additional hardware cost
\cite{wang}.

In \cite{5}, a distributed TDOA estimation method that relies only
on radio transceivers without other auxiliary measurement equipment
was presented. Ultra-wideband (UWB) signaling can be used to
accurately achieve time of arrival (TOA) or TDOA measurements, which
has the advantages of low-cost and penetrating ability, but also has
the weakness of short-range. TDOA based algorithms provide high
localization accuracy, and represent a practical method for
estimating range differences and source positions in WUSNs. However,
this method also faces many challenges. In particular, limited range
and directionality constraints decrease the accuracy of range
difference estimation. We notice that the TDOA can usually be
obtained from the measurement of a signal's phase which is
susceptible to phase ambiguity problems. The Chinese remainder
theorem (CRT) offers a closed-form analytical algorithm to calculate
a dividend from several of its corresponding divisors and
remainders, and can be applied to solve the ambiguity problem here.
In our ranging application, the remainders in the CRT are the
measured ``remainder'' wavelengths, the divisors are the measuring
wavelengths, and the dividend is the range difference to be
estimated. However, directly using the CRT is not feasible due to
its over-sensitivity to noise, i.e., a small error in a remainder
can lead to very large error in the estimated dividend. To avoid
this weakness of the CRT, we propose to use a robust Chinese
remainder theorem (RCRT) algorithm to estimate the range difference,
in which the dividend can still be reconstructed with only a small
error if the errors on remainders are bounded within a certain level
\cite{3}. As a result, the range differences or TDOAs can be
robustly estimated from noisy measurements in WUSNs by using the
RCRT.

After obtaining the range differences using the RCRT, we can
estimate the source position based on statistical signal processing
methods. For traditional terrestrial wireless sensor networks, TDOA
based localization algorithms are normally implemented in a
centralized way. In the centralized solution, all nodes relay their
TDOA measurements to a fusion center, which uses a conventional
localization algorithm to obtain the source position, and then sends
the global estimate back to every node. This strategy requires a
large amount of energy for communications \cite{6} and has a
potential failure point (the central node). Distributed strategies
are an attractive alternative, since they are in general more
robust, require less communications, and allow for parallel
processing. To address this limitation of centralized processing, we
propose distributed source localization algorithms using a diffusion
strategy in this paper. Diffusion algorithms were proposed in
\cite{6, 2010, 2007} and are applicable to distributed
implementations since nodes communicate in an isotropic manner with
their one-hop neighbor nodes, and no restrictive topology
constraints are imposed. Thus, the algorithms are easier to
implement and are also more robust to node and link failures. This
approach allows nodes to obtain better estimates than they would
without cooperation.

The rest of the paper is organized as follows. Section II
establishes the mathematical model of the problem, and derives the
proposed ranging method based on the RCRT. Section III gives the
distributed source localization algorithm based on the diffusion
strategy. Simulation results are given in Section IV, and
conclusions are drawn in Section V.

\section{System model and TDOA estimation via the robust CRT}

Due to the large attenuations in underground environments, experimental
results show that underground to underground (UG2UG) communication
is not feasible at the 2.4GHz frequency band \cite{10}. Underground
communication becomes practical only at lower frequencies. As reported
in \cite{11} and \cite{12}, a WUSN system operating at 433MHz with a
maximum transmit power of +10dBm usually achieves a communication
range of around one meter for UG2UG communication and more than 30m
for underground to aboveground (UG2AG) communication. These
communication ranges have already exceeded the wavelength of the
transmitted signal, which results in phase ambiguity  when the
distance difference is directly calculated from the phase. In this
section, we first propose a method based on the robust CRT to resolve
this phase ambiguity when computing the range distance.

\begin{figure}
  \includegraphics[width=2.5in]{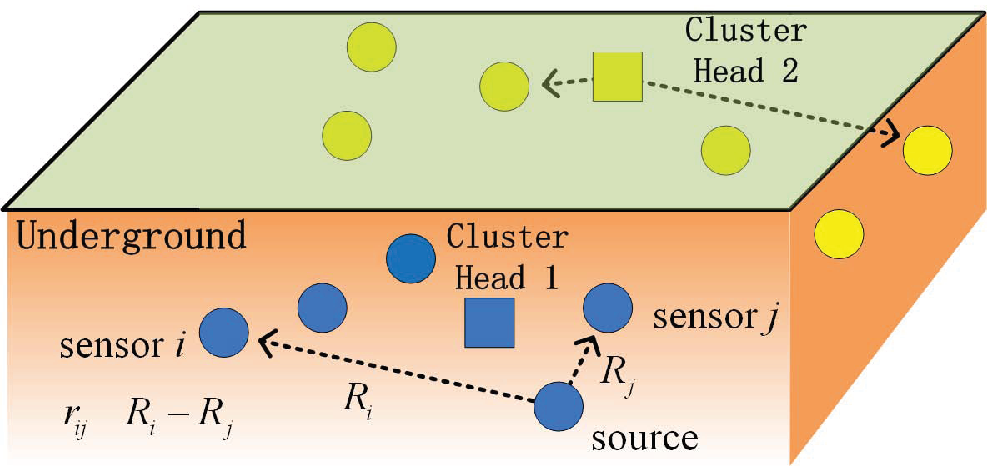}\\
  \caption{System model for WUSN localization.}\label{fig:WUSN}
\end{figure}

 Consider a WUSN with $L$ sensor nodes at known positions
\(({{x}_{i}},{{y}_{i}})\), \(i=1,2,\ldots ,L\), receiving a signal
from a source at an unknown position \((x^o,y^o)\) through a
dispersive medium, as shown in Fig. \ref{fig:WUSN}. The distance
between the source and the $i$-th sensor is
\begin{equation}\label{eq_Ri}
    {{R}_{i}}=\sqrt{{{({{x}_{i}}-x^o)}^{2}}+{{({{y}_{i}}-y^o)}^{2}}}.
\end{equation}
The range difference (RD) between the $i$-th and $j$-th receivers,
denoted by \({r_{ij}}\), is ${r_{ij}}={{R}_{i}}-{{R}_{j}}$.

If the medium is non-dispersive, the propagation delay for any
frequency is constant. However, in a dispersive medium, the signal
propagation velocity is a function of the signal's frequency,
denoted by $v_k$ for frequency ${\omega_k}$; that is, the
propagation delay is frequency dependent. We denote the delay of
propagation at frequency $\omega_k$ for sensor $i$ as
\begin{equation}\label{eq:t^k_i}
    {{\tau}_{k,i}}=\frac{{{R}_{i}}}{v_{k}}.
\end{equation}

We assume that the source transmits a sinusoid signal
${{s}_k}(t)=e^{j\, \omega_k t}$ at frequency $\omega_k$. The $i$-th
sensor receives the signal as
\begin{equation}
\begin{split}
 {{s}_{k,i}}(t)=e^{j\,\omega_k(t-\tau_{k,i})}+n_i(t),
\end{split}
\end{equation}
where $n_i(t)$ is the noise at sensor $i$, and \(\{{{n}_{i}}(t)\},
i=1,2,\ldots ,L\) are independent white Gaussian noise processes.
Similarly, the received signal at sensor $j$ is
\begin{equation}
s_{k,j}(t)=e^{j\, \omega_k (t-\tau_{k,j})}+n_j(t).
\end{equation}
By taking the cross-correlation of $s_{k,i}(t)$ and $s_{k,j}(t)$, we
get
\begin{equation}
I_{k,ij}= \frac{\int ^T _0 [ s_{k,i}(t) ]^* s_{k,j}(t) dt}{T}.
\end{equation}
It is easy to show that $I_{k,ij}$ is an asymptotically unbiased
estimator of $e^{j\omega_k(\tau_{k,i}-\tau_{k,j})}$
\cite{15:Proakis}, for which it follows that
\begin{equation}
\lim _{T\rightarrow \infty } \text{E}
\{I_{k,ij}\}=e^{j\omega_k(\tau_{k,i}-\tau_{k,j})}= e^{ j
\frac{{{\omega }_k}{r_{ij}}}{v_k} },
\end{equation}
where the range difference $r_{ij}$ is contained in the phase of
$I_{k,ij}$. We denote the phase of $I_{k,ij}$ by $\phi_{k,ij}$,
i.e.,
\begin{equation}
{\phi }_{k,ij}=\left( \frac{{{\omega }_k}{r_{ij}}}{v_k} \right)\bmod
2\pi.
\end{equation}
One can determine ${r_{ij}}$ from ${\phi }_{k,ij}$. However, there
are two issues of concern with this approach: 1) the value of ${\phi
}_{k,ij}$ is folded by $2\pi$ and 2) the measurements are noisy,
i.e.,
\begin{equation}
\frac{{{\omega }_k}{r_{ij}}}{v_k}={\phi }_{k,ij}+2\pi b_k+\nu_k,
\end{equation}
where $b_k$ is the quotient (folding integer) and $\nu_k$ is the
noise at frequency ${{\omega }_k}$.

For issue 1, since ${\phi }_{k,ij}\in[0, 2\pi)$, no matter how large
the actual ${r_{ij}}$ is, the RD ${r_{k,ij}}={\phi }_{k,ij}\cdot
v_k/{{\omega }_k}$ converted directly from ${\phi }_{k,ij}$ is
always within one wavelength of the signal, which is ${{\lambda
}_k}=2\pi v_k/{{\omega }_k}$. Therefore, $r_{ij}$ cannot be
determined uniquely from a single $\phi_{k,ij}$. We add more
constraints to confine the solution space by measuring the phase at
different frequencies, $\omega_k, k=1,2,\ldots,K$. The Chinese
remainder theorem (CRT) provides a solution to evaluate a dividend
from its remainders. We can regard ${r_{ij}}$ as the dividend,
${{\lambda }_k}$ as the divisor, and ${r_{k,ij}}$ as the
corresponding remainder, i.e.,
\begin{equation}\label{eq:remainder}
    {r_{ij}} = b_k {\lambda }_k  +{r_{k,ij}}.
\end{equation}
CRT tells that a positive integer $r_{ij}$ can be uniquely
reconstructed from its remainders $r_{k,ij}$ modulo $K$ positive
integers $\lambda_k$, if $r_{ij}<\text{lcm}\{\lambda_k\},k=1,...,K$,
where lcm$\{\cdot\}$ denotes least common multiple. It seems that
the CRT gives a perfect solution to this problem. However, when we
consider measurement noise, the traditional CRT
\cite{14:Koshy_Elementary} is not suitable because it is sensitive
to noise. A small error in the remainder can result in a large error
in the estimated dividend. Therefore, we adopt a robust CRT method
\cite{3}. Theorem 1 in \cite{3} proves that the robust CRT can
tolerate an error in ${r_{k,ij}}$ that is bounded by $\upsilon
<B/4$, where $B$ is the greatest common divisor (GCD) among the
divisors. To be more specific, if all the remainder errors are not
greater than the error bound $\upsilon$, the estimation error of the
unknown dividend is upper bounded by $\upsilon$, i.e.
\begin{equation}\label{eq:errbound}
\left | r_{ij}-\hat{r}_{ij}\right| \le \upsilon.
\end{equation}

Based on this robust CRT, we provide the following solution to solve
the problem:\\
\textbf{Step 1}. Estimate ${\phi }_{k,ij}$ by calculating the phase
of $I_{k,ij}$. Denote the estimated phase by ${\tilde{\phi
}}_{k,ij}\in [0,2\pi )$, and the corresponding distance converted
from the phase by
\begin{equation}\label{eq:tr_ij^k}
{{\tilde{r}_{k,ij}}}=\frac{{{{\tilde{\phi }}}_{k,ij}}}{2\pi
}{{\lambda }_k},\quad \text{where } {{\tilde{r}_{k,ij}}}\in
[0,{{\lambda }_k}), k=1,...,K.
\end{equation}
Note that the CRT is commonly expressed over the ring of integers
while the distance is a real number. Therefore, we extend the
algorithm from the integers to the reals by introducing a real
common factor among the divisors as in \cite{16:Wang}. We choose
those communication frequencies so that the ${{\lambda }_k},
k=1,...,K$, have a real common factor of $B$ which satisfies
\begin{equation}\label{eq:lambda^k}
{{\lambda }_k}=B{\Gamma }_k,
\end{equation}
where ${{\Gamma }_k}$ are co-prime integers, i.e., $\left( {{\Gamma
}_m},{{\Gamma }_{n}} \right)=1$, for $1\le m,n\le K,m\ne n$, and
$\left( \cdot , \cdot  \right)$ denotes GCD. According to the prior
discussion,  we have the following equation:
\begin{equation}
\begin{split}
 {r_{ij}}&= {{b}_k}{{\lambda }_k}+{{r}_{k,ij}} \,\,\,\, ,k=1,...,K,\\
 &= {{b}_{k}}{{\lambda }_{k}}+{{{\tilde{r}}}_{k,ij}}+\Delta
 {{r}_{k,ij}}
 \end{split}
\end{equation}
where $\Delta {{r}_{k,ij}}$ denotes the measurement errors of
${{\tilde{r}}_{k,ij}}$ with $\left| \Delta {{r}_{ij}^k}
\right|<\upsilon
$, and $\upsilon$ is the error bound.\\
 \textbf{Step 2}. For notational convenience, define the following
 symbols:
\begin{equation}
\Gamma \triangleq \prod\limits_{k=1}^{K}{{{\Gamma }_k}},
\end{equation}
\begin{equation}
{{\gamma }_{k}}=\Gamma /{{\Gamma }_k}.
\end{equation}
Calculate and find the sets,
\begin{equation}\label{13}
{{S}_{k}}\triangleq \left\{
({{{\bar{b}}}_{1}},{{{\bar{b}}}_{k}})=\arg
\underset{({{{\hat{b}}}_{1}},{{{\hat{b}}}_{k}})\in {{\Omega
}_{k}}}{\mathop{\min }}\,\left| {{{\hat{b}}}_{k}}{{\lambda
}_{k}}+{{{\tilde{r}}}_{k,ij}}-{{{\hat{b}}}_{1}}{{\lambda
}_{1}}-{{{\tilde{r}}}_{1,ij}} \right| \right\},k=2,3,...,K,
\end{equation}
where
\begin{equation}
{{\Omega }_k}\triangleq \left\{
({{{\hat{b}}}_{1}},{{{\hat{b}}}_{k}})|0\le {{{\hat{b}}}_{1}}\le
{{\gamma }_{1}}-1,0\le {{{\hat{b}}}_{k}}\le {{\gamma }_{k}}-1
\right\}
\end{equation}
is the solution space of the quotients.

The set $S_k$ can be regarded as the optimal combination of the
quotients $b_1$ and $b_k$ with which the difference of the estimated
dividends achieves its minimum.
\\
\textbf{Step 3}. Let \(S_{k,1}\) denote the first element
\({{\bar{b}}_{1}}\) of the 2-tuples,
\begin{equation}
S_{k,1}\triangleq
\{{{\bar{b}}_{1}}|({{\bar{b}}_{1}},{{\bar{b}}_{k}})\in {{S}_{k}}\
\textrm{for some} \ {{\bar{b}}_{k}}\}.
\end{equation}
Calculate the intersection set of \({{S}_{k,1}}\):
\begin{equation}
S\triangleq \bigcap\limits_{k=2}^{K}{{{S}_{k,1}}}.
\end{equation}
In \cite{3}, it was proved that if the error bound is less than
\(B/4\), the set \(S\) contains only the true value of
\({{b}_{1}}\), i.e., \(S=\{{{b}_{1}}\}\). In addition, $b_k$ can
also be determined from $S_k$, that is, if
$({{b}_{1}},{{\bar{b}}_{k}})\in {{S}_{k}}$, then
${{\bar{b}}_{k}}={{b}_{k}}$ for \(2\le k\le K\). Therefore, with all
the quotients being determined correctly, the error of the estimated
\({r_{ij}}\) is therefore bounded by (\ref{eq:errbound}), where
$\hat{r}_{ij}$ is obtained by averaging the estimates corresponding
to different wavelengths, i.e.,
\begin{equation}
{{\hat{r
}}_{ij}}=\frac{1}{K}\sum\limits_{k=1}^{K}{({{b}_{k}}{{\lambda
}_{k}}+{{{\tilde{r}}}_{k,ij}})}.
\end{equation}

\emph{Remark:} One may notice that to find the set $S_k$ in
(\ref{13}) we need to search among the possible values of
$\Omega_k$, which is a 2-D search problem with the order of
$(\Gamma_2 \Gamma_3 ... \Gamma_K)^2$ and requires a high
computational complexity. However, the complexity can be decreased
by using a fast algorithm proposed in \cite{3} and
\cite{13:Li07efficient} to a 1-D search problem with the order of
only $2(L-1)\Gamma_i$. One can easily apply the fast algorithm to
the application in this paper. We do not describe this fast
algorithm herein since this is not the focus of this paper.

\section{Diffusion Algorithms for Source Localization}

After obtaining the range differences using the RCRT, a localization
algorithm can be used to estimate the source position. Herein, we
propose an algorithm for source localization that employs diffusion
strategies proposed in \cite{6, 2010, 2007}. First, the nodes in a
cluster send their measurements to the cluster head. Then, the
cluster head determines a local estimate using these measurements
locally. After obtaining the local estimates, cluster heads exchange
their local estimates to achieve diffusion. Before describing the
distributed diffusion process, we first discuss the global WLS
estimate when all measurements are sent to a fusion center for
estimating the source location.

\subsection{Global Weighted Least Squares Problem}

Consider a set of $N$ cluster heads and $K = MN$ sensor nodes (each cluster head is associated with $M$ sensor nodes) spatially distributed over some region with known locations $x_i$'s (we consider the 2-D Cartesian coordinate system).
The objective of the network is to collectively estimate an unknown deterministic column vector - the location $x$ of a source.
All the nodes (cluster heads and sensor nodes) in the network can measure the signal transmitted from the source.
The sensor nodes transmit their measurements to the corresponding cluster head and each cluster head forms a set of $M$ TDOA measurements with itself being the reference. Then, cluster heads can send their TDOA
measurements to the fusion center. The TDOA measurements are formed one at a time by comparing the signal from the cluster head and the signal from a sensor node, thus leading to uncorrelated estimates if the estimation period is longer than the typical coherence time of the mobile radio channel. Finally, the fusion center obtains altogether $K$ TDOA based range difference measurements $r_{i,j}$'s in the network.

The scalar model for TDOA based range difference is given by
\begin{equation} \label{model}
r_{i,j} = \|x-x_i\| -  \|x-x_j\| + n_{i,j},
\end{equation}
stack all the $r_{i,j}$'s into a vector and write
\[
r = s(x) + n
\]
where $t = \mathrm{col}\{ r_{i,j}\}$, $s(x) = \mathrm{col}\{\|x-x_i\| -  \|x-x_j\|\}$ and $n = \mathrm{col}\{n_{i,j}\}$. Each $r_{i,j}$ can be alternatively denoted as $[r]_l$ to reflect its location in vector $r$.
The corresponding covariance matrix for $n$ is denoted as $W$ and $W = \mathrm{diag}\{\sigma_1^2, \dots, \sigma_K^2\}$.

The global weighted least squares estimator for estimating $x$ given $t$ is thus
\begin{eqnarray} \label{global}
&&\hat{x}_G = \arg \min (r- s(x))^T W^{-1} (r-s(x)) \nonumber\\
&&\ \ \ \ = \arg \min \sum_{i=1}^K \sigma_i^2 ([r]_i - [s(x)_i])^2.
\end{eqnarray}

Assuming the $n_{i,j}$'s are Gaussian, then the covariance of $\hat{x}_G$ attains the corresponding CRLB, which is given by \cite{13}
\begin{equation} \label{crlb}
\mathrm{cov}(\hat{x}_G) = (P^T W^{-1} P)^{-1}
\end{equation}
where
\begin{equation}
[P]_{i,j} = \frac{\partial [s(x)]_i}{\partial [x]_j}\Big|_ {x = x_0}
\end{equation}
and $x_0$ denotes the true position of the source.

For the TDOA measurements, $P$ is a $K \times 2$ matrix and the elements of $P$ are given by
\begin{eqnarray}
&&[P]_{l,1} = \frac{[x_i]_1 - [x]_1}{\|x_i - x\|} - \frac{[x_j]_1 - [x]_1}{\|x_j - x\|}\Big|_ {x = x_0}, \nonumber \\
&&[P]_{l,2} = \frac{[x_i]_2 - [x]_2}{\|x_i - x\|} - \frac{[x_j]_2 - [x]_2}{\|x_j - x\|}\Big|_ {x = x_0} \nonumber
\end{eqnarray}
where we assume $[s(x)]_l$ involves nodes (a sensor node and a cluster head) $i$ and $j$.

\subsection{Local Weighted Least Squares Problem}

For each cluster head $k$, it has access to limited data from its
neighbors. It can then solve the WLS problem locally as
\begin{equation}
\hat{x}_k = \arg \min \sum_{i=1}^K c_{i,k} \sigma_i^2 ([r]_i - [s(x)_i])^2
\end{equation}
where $c_{i,k}$'s are the associated weights for node $k$ and
$c_{i,k} =0$ if $[t]_i$ is not accessible by node $k$. Let $C$
denote the $K \times N$ matrix with elements $c_{i,k}$. We require
$\mathds{1}^TC=\mathds{1}^T$, where $\mathds{1}$ denotes the $N
\times 1$ column vector with unit entries.

A local estimate can also be written as
\begin{equation} \label{local}
\hat{x}_k = (P^T W^{-1} C_k P)^{-1} P^T W^{-1} C_k z \triangleq L_k z
\end{equation}
where $C_k = \mathrm{diag}\{Ce_k\}$ ($e_k$ is an $N \times 1$ vector with a unity entry in position $k$ and zeros elsewhere) and $z = r - s(x) + P x |_{x = x_0}$. $P$ can be estimated at $\hat{x}_k$.

The covariance matrix associated with $\hat{x}_k$ is given by $\mathrm{cov}(\hat{x}_k) = L_k W L_k^T$. Here $L_k$ contains only local information due to the selection ability of $C_k$.

The estimation fusion method can then be used to fuse these local estimates in a distributed manner by utilizing the local covariance matrices as proper weights and the estimation fusion method can be shown to achieve the performance of the global estimation. However, it involves covariance estimation and matrix inversion.

\subsection{Diffusion Algorithm}

Besides estimator fusion, we can also use a diffusion algorithm to
perform distributed estimation. For each cluster head $k$, at the
$i$th time epoch, it exchanges its local estimate with its
neighboring cluster heads and updates its local estimate using a
diffusion algorithm:
\begin{equation}\label{conf10}
\hat{x}_{k,i} = \sum_{l=1}^N a_{l,k} \hat{x}_{l,i},
\end{equation}
where the $a_{l,k}$'s are the diffusion coefficients.
 Eq. (\ref{conf10}) can be considered as a weighted average of the local estimates in the neighborhood of node $k$. Assume all the local estimates are unbiased. Then in order for $\hat{x}_{k,i} $ to be unbiased, we require $\mathds{1}^T a_k=\mathds{1}^T$ where $a_k = [a_{1,k}, \dots, a_{N,k}]^T$. The diffusion process is repeated until all the local estimates have converged, i.e., $\|\hat{x}_{k,i+1} - \hat{x}_{k,i}\| \leq \epsilon \ \forall k$, where $\epsilon$ is a (small positive) design parameter.

One possible choice for the weights $a_{l,k}$ is to consider the degree of connectivity, which is
\begin{equation}\label{con}
a_{l,k} = \left\{
\begin{split}
&\mathrm{deg}_l / \sum_{n \in N_k} \mathrm{deg}_n, &l \in N_k\\
&0, &\textrm{otherwise}
\end{split}
\right.
\end{equation}
where $\mathrm{deg}_l$ denotes the cardinality of cluster head $l$'s neighbors (also cluster heads) and $N_k$ denotes the set of neighboring cluster heads of head $l$. Such choice has been observed to yield good results for the diffusion algorithm in general \cite{6}.

However, this method does not consider the reliability of different local estimates. The reliability of a local estimate is reflected in its associated covariance matrix. But estimating the covariance matrix is not an easy task for a nonlinear weighted least-squares estimator (WLSE). Here we propose a method for setting appropriate $a_{l,k}$'s which reflects the reliability of different local estimates to a certain extent without requiring use of the covariance matrices.

Since each local estimate contains certain errors, when sorting them in respective dimensions (each dimension of $\hat{x}_k$ is treated separately), the middle ones are more reliable.
Therefore, for cluster head $k$, at the $i$th time epoch, it first finds the median $\tilde{x}_{k,i}$ along respective dimensions among its received local estimates $\hat{x}_{l,i}$'s.
Then for each obtained local estimate, head $k$ calculates $w^i_{l,k} = \exp(-\|\hat{x}_{l,i} - \tilde{x}_{k,i}\|^2/\gamma)$ if $l \in N_k$. Otherwise, $w^i_{l,k}$ = 0. $\gamma$ is a parameter which controls how rapidly the weight decays as $\|\hat{x}_{l,i} - \tilde{x}_{k,i}\|$ grows.
The diffusion coefficient $a^i_{l,k}$ is then set to
\begin{equation}\label{weight}
a^i_{l,k} = w^i_{l,k} / \sum_{j \in N_k}  w^i_{j,k}.
\end{equation}
Here we explicitly indicate the time epoch of $w^i_{l,k}$ and $a^i_{l,k}$ in the superscript.
It can be seen that the larger the deviation between a local estimate $\hat{x}_{l,i}$ and $\tilde{x}_{k,i}$, the smaller the weight assigned to this local estimate and vice versa. Obviously, $\mathds{1}^T a^i_{l,k} = \mathds{1}^T$.

Another method of interest is to use an optimization technique. That is, to set $a_{l,k}$'s such that the trace of the covariance matrix of $\hat{x}_{k,i}$ is minimized. We start by examining the first time epoch $i=1$. Then, we have
\begin{equation}
\hat{x}_{k,i} = \sum_{l=1}^N a^i_{l,k} \hat{x}_{l,i} = \sum_{l=1}^N a^i_{l,k} L_{l,i} z
\end{equation}
where we have used (\ref{local}).

The covariance matrix of $\hat{x}_{k,i}$ is thus given by
\begin{equation}
\mathrm{cov}(\hat{x}_{k,i}) = (\sum_{l=1}^N a^i_{l,k} L_{l,i}) W (\sum_{l=1}^N a^i_{l,k} L_{l,i})^T
\end{equation}
and its trace is
\begin{equation}
\mathrm{tr}(\mathrm{cov}(\hat{x}_{k,i})) = \sum_{m,n=1}^N a^i_{m,k} a^i_{n,k} \mathrm{tr}(W L_{m,i}^T L_{n,i})
= (a^i_k)^T Q a^i_k
\end{equation}
where $a^i_k = [a^i_{1,k}, \dots, a^i_{N,k}]^T$ and $[Q]_{m,n} = \mathrm{tr}(W L_{m,i}^T L_{n,i})$.
To be unbiased, we also require $\mathds{1}^T a^i_k = \mathds{1}^T$.

Therefore, to find the optimal $a^i_k$ in the sense of minimizing the diffusion covariance matrix, we need to solve an optimization problem:
\begin{equation}\label{opt}
\hat{a}^i_k = \arg \min (a^i_k)^T Q a^i_k, \ s.t. \ \mathds{1}^T a^i_k = \mathds{1}^T, \ a^i_k\geq 0, \ a^i_{l,k} = 0 \ \forall \  l \notin N_k.
\end{equation}
This problem is convex if $Q \succ 0$ and thus can be solved fast and efficiently.

After determining $\hat{a}_k^i$ at time epoch $i$, $L_{l,i+1}$ is updated as $L_{l,i+1} = \sum_{l=1}^N \hat{a}^i_{l,k} L_{l,i}$ and $Q$ will also be updated accordingly. Then the above optimization process can be performed iteratively until estimates converge.

This optimization method can enhance the distributed fusion performance at the cost of slightly higher computational complexity compared with simply setting $a_{l,k}$ according to (\ref{con}).

To decrease the computational cost, the optimization process can be implemented only once for the first diffusion at each node, and the weight $a_k$ remains the same for the latter diffusions. Since after one diffusion, the updated local estimates becomes much less dissimilar and thus the weights will have much less influence on the followed diffusions.

\section{Simulation results}
In this section, simulation results are given. We first study the ranging performance using the RCRT in Subsection \emph{A}.
Then, the localization accuracy is introduced in Subsection \emph{B}.
\subsection{Ranging Performance}
\begin{figure}[!t]\label{fig:1}
\centering
\includegraphics[width=3.3in]{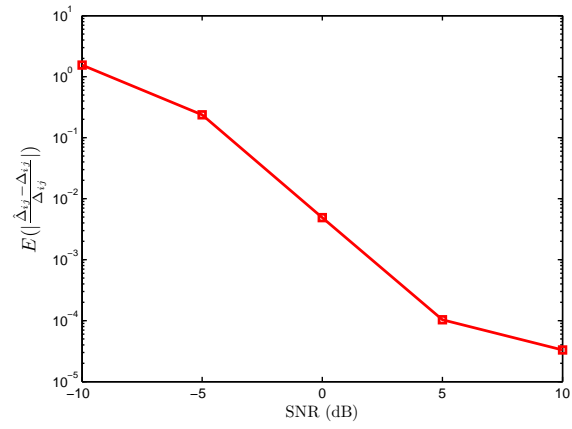}
\caption{Relative estimation error decreases with increasing
SNR.}
\end{figure}

Assume that the signals are transmitted at 3 frequencies, i.e.,
\(K\)=3, \(B\)=80, \({{\Gamma }_{k}}\)=\(\{15,16,17\}\), and the
corresponding three dividers are \(\{1200,1280,1360\}\) which
represent the wavelengths ${{\lambda
}_{k}}=\{120,128,136\}\text{mm}$. According to the RCRT, the maximum
estimation distance is \({{d}_{\max
}}=B\prod\limits_{k=1}^{K}{{{\Gamma }_{k}}}\)=32640mm. Each trial of
simulation generates random integers \({r_{ij}}\), which are
uniformly distributed over \([0,{{d}_{\max }}]\). And there are 1000
trials for each SNR. The result is shown in Fig. \ref{fig:1}.

Then we consider the effect of \(B\) on the performance. We set
\(B=100,200,300\), respectively and fix \({{\Gamma }_{k}}\)=\(\text{
}\!\!\{\!\!\text{ 7,9 }\!\!\}\!\!\text{ }\). The result is shown in
Fig. 2 with 10000 trials for each SNR. According to the result,
changing \(B\) does not have significant influence on the relative
error of distance estimation.

Next, we compare the performance under different values of
$\Gamma_k$ with constant \(B\) and \(K\). In the simulation, we fix
\(B\)=50, \(K\)=3, and let $\Gamma_k$ be \(\{7,11,15\}\),
\(\{29,33,37\}\), \(\{53,57,61\}\), respectively. Fig.3 demonstrates
that the estimation error increases with \({{\Gamma }_{k}}\). The
results from Figs. 2 and 3 can be explained by equations
(\ref{eq:tr_ij^k}) and (\ref{eq:lambda^k}): the phase measurement
error is amplified by \(B\) and $\Gamma_k$, i.e., the error of
${{\tilde{r}}_{k,ij}}$ is \(\Delta {{r}_{k,ij}}=\frac{\Delta
{{{\tilde{\phi }}}_{k,ij}}}{2\pi }B{{\Gamma }_{k}}\), where \(\Delta
{{\tilde{\phi }}_{k,ij}}\) denotes the phase measurement error of
\({{\tilde{\phi }}_{k,ij}}\). The larger $\Gamma_k$ is, the larger
error results. However, $B$ does not affect the performance because
the robustness of the algorithm also linearly increases with $B$
which cancels the performance deterioration of the increased $B$.
(The error tolerance of the algorithm is $\upsilon <B/4$)

In addition, we consider the scenario in which different sets of
$\Gamma_k$ are compared under the constraint of constant maximum
range \({{d}_{\max }}=B\prod\limits_{k=1}^{K}{{{\Gamma }_{k}}}\). In
Fig. 4, We choose \({{\Gamma }_{k}}\)=\(\{7,11,15\}\),
\(\{5,11,21\}\), \(\{3,11,35\}\), respectively. The simulation
results suggest that the performance is better if the differences
between the \({{\Gamma }_{k}}\) are smaller.

Fig. 5 demonstrates that using more wavelengths results in better
ranging performance. In this simulation, we fix the maximum
estimation distance \({{d}_{\max }}\), and vary $K$. We consider the
cases when \(K\)=\(\text{2,3,4}\), respectively, with \(B\)=50 and
\({{d}_{\max }}\)=3465mm. \({{\Gamma }_{k}}\) are set to \(\text{
}\!\!\{\!\!\text{ }33,35\text{ }\!\!\}\!\!\text{ }\), \(\text{
}\!\!\{\!\!\text{ }7,11,15\text{ }\!\!\}\!\!\text{ }\), \(\text{
}\!\!\{\!\!\text{ }3,5,7,11\text{ }\!\!\}\!\!\text{ }\),
respectively. Simulation results demonstrate that our RCRT based
ranging scheme can estimate the range differences with high
accuracy.

\subsection{Comparison Results for Distributed Source Localization}

We now show simulation results for source localization.
The algorithms compared are: 1) \textit{Global}: the global weighted least square estimator (\ref{global}), 2) \textit{Diff (con)}: the diffusion algorithm with coefficients set by considering connectivity (\ref{con}), 3) \textit{Diff (wei)}: the diffusion algorithm with coefficients set by weighting (\ref{weight}), 4) \textit{Diff (opt)}: the diffusion algorithm with coefficients set by optimization (\ref{opt}) and 5) \textit{Local}: simple average of all the local estimates, i.e., $\sum_{k = 1}^N \hat{x}_k/N$.

The root mean square error (RMSE) is used as a performance metric, which is defined as $\sqrt{E (\| \hat{x} - x\|^2)}$.
The Cramer-Rao Lower Bound (CRLB) on the RMSE based on the entire data (equals to $\sqrt{\mathrm{tr}(\mathrm{cov}(\hat{x}_G))}$ for Gaussian measurement noise) is also presented as a benchmark.
Each simulated point is averaged over 200 runs.

The simulated network consists of $N$ cluster heads that are regularly deployed at grid points. The distance between neighboring cluster heads is set to 50 (the units are meters, here and below). Each cluster head has $M$ associated sensor nodes which are distributed uniformly around the corresponding cluster head. The source location is fixed at $[60,70]$ in all the simulations. The TDOA measurements are generated according to (\ref{model}) with $n_{i,j}$'s being Gaussian noises. $W$ is set to $W = \sigma^2 I$ with $\sigma=1$. $\gamma$ is set to 1. The initial point for the WLSE is always set as the center of the deployment area.

Here we consider the scenario in which each cluster head exchanges its local measurements with its neighboring cluster heads to perform a local estimate, and then exchanges its local estimate with its neighboring cluster heads to perform diffusion until convergence. $C$ is set to
\begin{equation}
c_{i,k} = \left\{
\begin{split}
&\check{c}_{l,k}, &[t]_i \textrm{ involves cluster head } l, l \in N_k \textrm{ and } l \neq k\\
&\check{c}_{k,k}, &[t]_i \textrm{ involves cluster head } k\\
&0, &\textrm{otherwise}
\end{split}
\right.
\end{equation}
where
\begin{equation}
\check{c}_{l,k} = \left\{
\begin{split}
&1/ \max\{\mathrm{deg}_l, \mathrm{deg}_k\}, &l \in N_k, \ l \neq k\\
&1- \sum_{l \neq k} \check{c}_{l,k},  &l = k\\
&0, &\textrm{otherwise}.
\end{split}
\right.
\end{equation}
Here, $\check{c}_{l,k}$ denotes the weight assigned with respect to cluster heads $l$ and $k$, and $c_{i,k}$ represents the weight assigned to the $i$th measurement used by the cluster head $k$.

First, we fix $N =16$ and vary the value of $M$, which changes from 5 to 50 with a step size of 5. The RMSEs of respective algorithms are shown in Fig. \ref{rmse}. It can be observed that \textit{Global} is the best which can attain the CRLB, \textit{local} is the worst and \textit{Diff (con)} is always better than \textit{local}. In general, \textit{Diff (opt)} is better than \textit{Diff (wei)}, and \textit{Diff (wei)} is better than \textit{Diff (con)}. The performance improvement of \textit{Diff (opt)} and \textit{Diff (wei)} compared with \textit{Diff (con)} comes from the consideration of the reliability of the local estimates. Though the diffusion algorithms are always worse than \textit{Global}, the performance differences are not significant. As $M$ grows, all the algorithms perform better.

Fig. \ref{time} shows the corresponding average CPU times of respective algorithms except \textit{Diff (opt)} whose CPU time is typically 10 times that of \textit{Global} due to its numerical optimization nature (the same below). It can be seen that \textit{Diff (con)} is very time efficient with a CPU time almost the same as \textit{local}. \textit{Diff (wei)} consumes a little more CPU time than \textit{Diff (con)}, while the CPU time of \textit{Global} is much larger. This demonstrates the advantage of the diffusion algorithm in terms of CPU time and computational complexity. Furthermore, we can say the diffusion algorithm has lower communication cost and computation complexity than the centralized solution, while the localization accuracy is close to the centralized method.

Fig. \ref{niter} shows the average number of iterations before convergence of the three diffusion algorithms. We can observe that \textit{Diff (opt)} requires many fewer iterations compared with \textit{Diff (con)} and \textit{Diff (wei)}. \textit{Diff (wei)} needs slightly more iterations than \textit{Diff (con)}, which may explain our observation of a slightly longer CPU time consumed by \textit{Diff (wei)}.

We now fix $M = 10$ and examine the effects of the number of cluster heads $N$ on the performance of the algorithms. We vary $N$ from 4 to 36. The RMSEs of the algorithms are shown in Fig. \ref{rmse1}. It is clear that adding more clusters (and thus sensor nodes) will not necessarily improve the estimation performance. This is known as the geometric effect of the localization problem. Since the source location is outside the convex hull formed by respective added clusters, the corresponding local estimates are not good enough. \textit{Diff (con)} and \textit{Diff (wei)} thus also show performance degradation. However, \textit{Diff (wei)} is much better than \textit{Diff (con)} when more clusters are added while the sensor nodes associated with each cluster head is fixed. The performance of \textit{Global} is almost unchanged as $N$ becomes large. Due to the consideration of the estimation covariance matrices, \textit{Diff (opt)} shows very good performance. However, it will consume much more CPU time. The corresponding average CPU times and average number of  iterations are shown in Figs. \ref{time1} and \ref{niter1} respectively. Similarly, \textit{Diff (con)} is the most time efficient and \textit{Diff (opt)} requires the smallest number of iterations.

Then we examine the effect of $\sigma$ on the performance of the algorithms. We set $N = 16$ and $M=10$. The RMSEs of respective algorithms are shown in Fig. \ref{rmse2}. The relative performance of respective algorithms is the same as before. As $\sigma$ enlarges, all of them show performance degradation. The average CPU times and average number of  iterations among different algorithms have the same relationship as before and thus the figures are not shown here.

Finally, we examine the choice of $\gamma$ on the performance of \textit{Diff (wei)}. We set $N = 16$, $M=10$ and $\sigma =1$. We generate 200 realizations of the overall TDOA measurement vector, then store and use them for all the corresponding simulations. The corresponding RMSEs are shown in Fig. \ref{rmse3}. It can be seen that an optimal $\gamma$ exists which results in the minimum RMSE for \textit{Diff (wei)}. However, in a large range of the choice of $\gamma$, \textit{Diff (wei)} can generate better results than \textit{Diff (con)}.

\section{Conclusion}
We have presented energy efficient localization schemes that can achieve high localization accuracy in
wireless underground sensor networks. These distributed localization algorithms require low computational complexity and energy consumption based on a diffusion strategy. An accurate RCRT based ranging scheme using TDOA to determine range differences between sensors and source that does not require time synchronization is also proposed. It has been shown via simulation results that the proposed localization algorithms achieve excellent localization accuracy with lower communication cost. In future work, we plan to implement our localization scheme in a testbed and verify its performance with an actual WUSN.

\newpage

\begin{figure}[!t]
\centering
\includegraphics[width=3.3in]{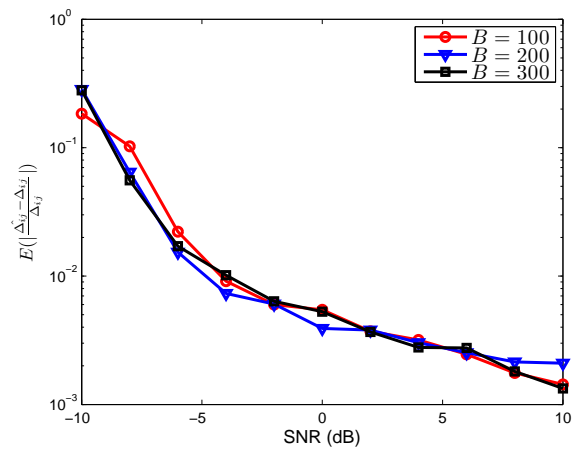}
\caption{Comparison between the relative estimation errors for
different values of \(B\).}
\end{figure}

\begin{figure}[!t]
\centering
\includegraphics[width=3.3in]{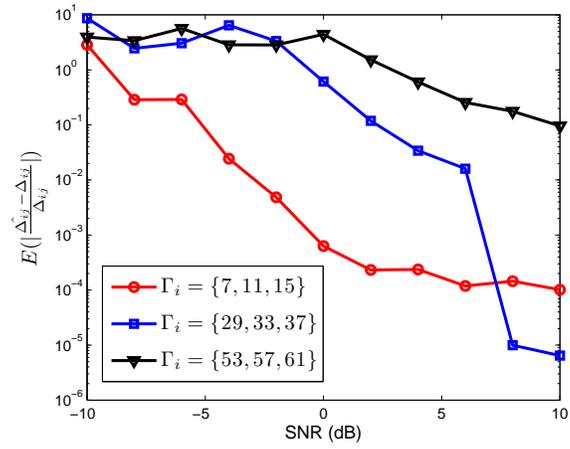}
\caption{Comparison of the relative estimation error for different
values of \({{\Gamma }_{k}}\), when \(B\) and \(K\) are fixed.}
\end{figure}

\begin{figure}[!t]
\centering
\includegraphics[width=3.3in]{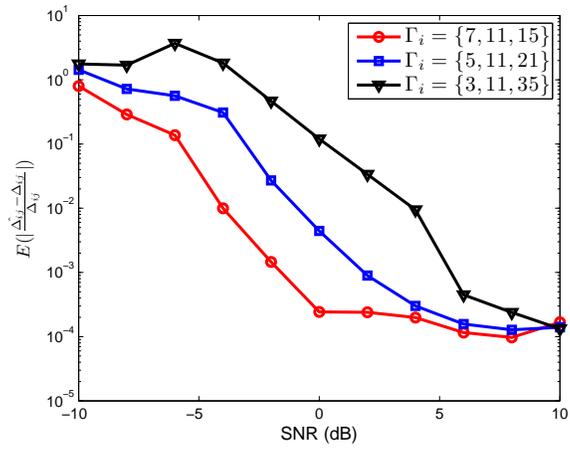}
\caption{Comparison of the relative estimation error for different
values of \({{\Gamma }^{k}}\), with \(B\), \(K\) and \({{d}_{\max
}}\) fixed.}
\end{figure}

\begin{figure}[!t]
\centering
\includegraphics[width=3.3in]{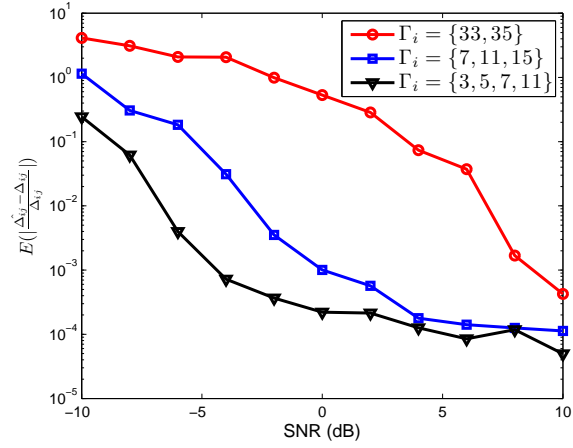}
\caption{Comparison of the relative estimation error for
different values of \(K\).}
\end{figure}

\begin{figure}[!t]
\centering
\includegraphics[width=0.48\textwidth]{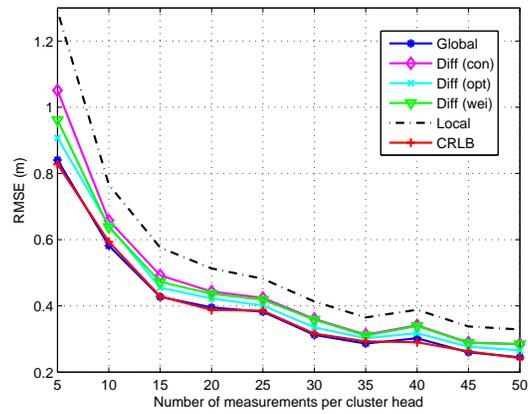}
\caption{RMSE versus $M$ when $N = 16$.}
\label{rmse}
\end{figure}

\begin{figure}[!t]
\centering
\includegraphics[width=0.48\textwidth]{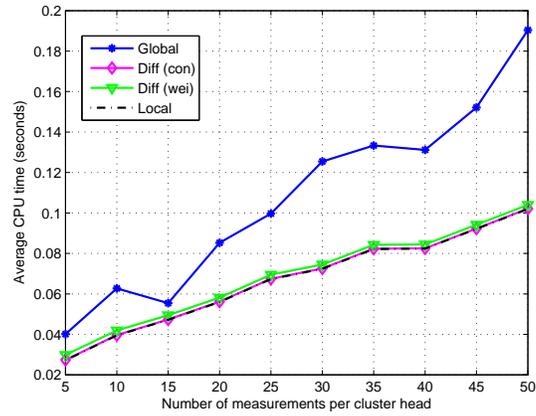}
\caption{CPU time versus $M$ when $N = 16$.}
\label{time}
\end{figure}

\begin{figure}[!t]
\centering
\includegraphics[width=0.48\textwidth]{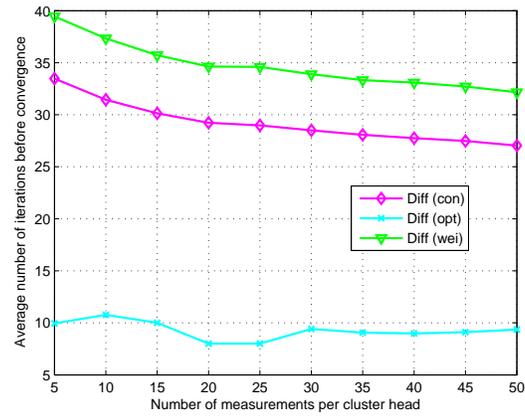}
\caption{Average number of iterations before convergence versus $M$ when $N = 16$.}
\label{niter}
\end{figure}

\begin{figure}[!t]
\centering
\includegraphics[width=0.48\textwidth]{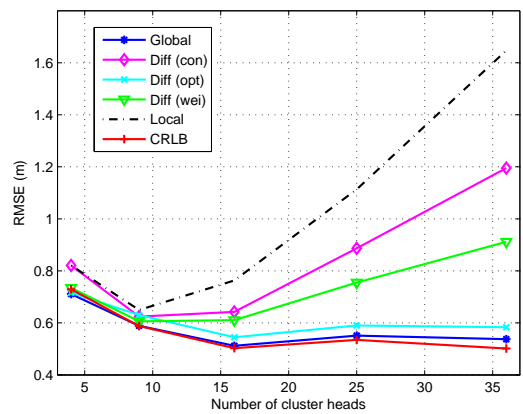}
\caption{RMSE versus $N$ when $M = 10$.}
\label{rmse1}
\end{figure}

\begin{figure}[!t]
\centering
\includegraphics[width=0.48\textwidth]{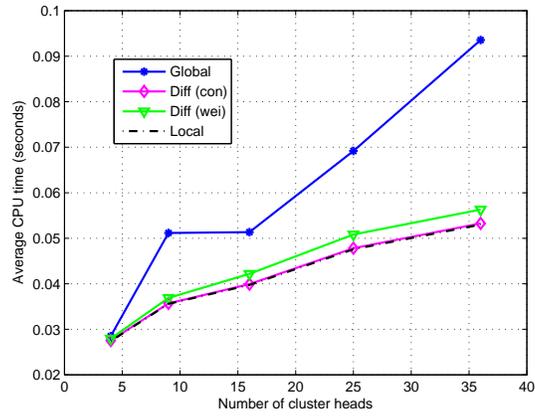}
\caption{CPU time versus $N$ when $M = 10$.}
\label{time1}
\end{figure}

\begin{figure}[!t]
\centering
\includegraphics[width=0.48\textwidth]{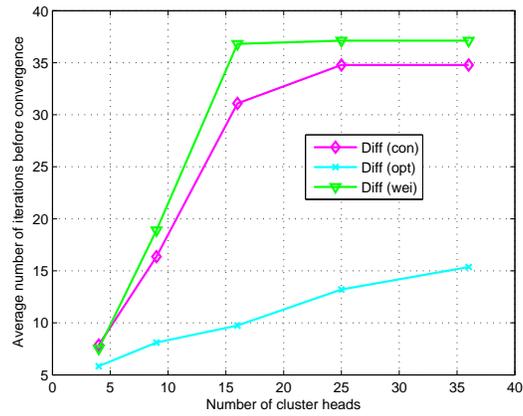}
\caption{Average number of iterations before convergence versus $N$ when when $M = 10$.}
\label{niter1}
\end{figure}

\begin{figure}[!t]
\centering
\includegraphics[width=0.48\textwidth]{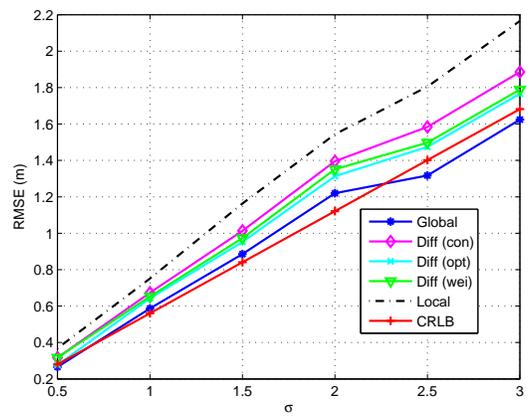}
\caption{RMSE versus $\sigma$.}
\label{rmse2}
\end{figure}

\begin{figure}[!t]
\centering
\includegraphics[width=0.48\textwidth]{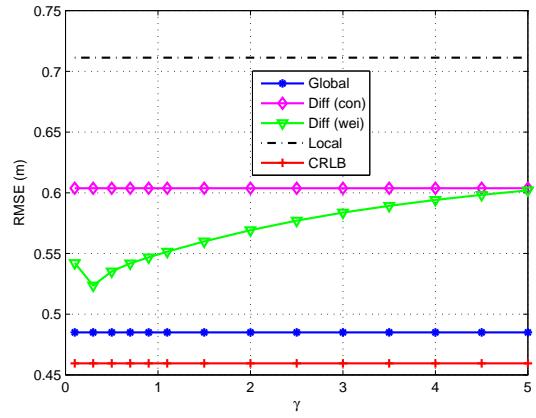}
\caption{RMSE versus $\gamma$.}
\label{rmse3}
\end{figure}

\end{document}